\documentclass[aps,prd,reprint,amsmath,amssymb,superscriptaddress,floatfix]{revtex4}
\usepackage{graphicx}
\usepackage{verbatim} 
\usepackage{amsfonts}
\usepackage{amssymb}
\usepackage{rotating}
\usepackage{booktabs}
\usepackage{xcolor}
\usepackage{soul}
\usepackage{color}
\usepackage{slashed}
\usepackage{multirow}
\usepackage{makecell}
\usepackage{epsf}
\usepackage{ulem}
\usepackage{cancel}
\usepackage{color,bm}
\usepackage[colorinlistoftodos]{todonotes}
\usepackage{diagbox}
\usepackage[colorlinks=true,citecolor=cyan,urlcolor=blue,bookmarks=true,bookmarks=true,bookmarksopen=true,bookmarksnumbered=true,bookmarksopenlevel=3]{hyperref}
\usepackage{subfigure}
\definecolor{airforceblue}{rgb}{0.36, 0.54, 0.66}
\definecolor{steelblue}{rgb}{0.27, 0.51, 0.71}
\definecolor{amber}{rgb}{1.0, 0.49, 0.0}

\begin{document}

\title{Dark matter, leptogenesis and $Z^\prime$ in the $B-L$ model}

\author{\textsc{XinXin Qi}}
\email{qxx@dlut.edu.cn}
\affiliation{Institute of Theoretical Physics, School of Physics, Dalian University of Technology, No.2 Linggong Road, Dalian, Liaoning, 116024, P.R.China }
\author{\textsc{Hao Sun}}
\email{haosun@dlut.edu.cn}
\affiliation{Institute of Theoretical Physics, School of Physics, Dalian University of Technology, No.2 Linggong Road, Dalian, Liaoning, 116024, P.R.China }

\begin{abstract}
We discuss the interplay between dark matter, leptogenesis and a new gauge boson $Z^\prime$ in the $B-L$ model. A fermion dark matter $\chi$ carrying $U(1)_{B-L}$ charge is introduced to the model but not coupling with other particles. We consider the freeze-out and freeze-in mechanisms, and obtain the correct relic density respectively. 
We have scanned the feasible parameter space and found that the dark matter direct detection experiments imposed the most stringent constraints on the parameter space. 
The constraint on the parameter space places a limit on the mass of dark matter with $m_{\chi} \approx 1/2 M_{Z'}$ within a narrow region in the case of freeze-out scenario, 
and we can obtain the right baryon asymmetry result in the case of  $m_{\chi}\subset[654$ GeV, 664 GeV]. 
For the freeze-in scenario, we have a much broader parameter space for $m_{\chi}$ and $M_{Z^\prime}$ but $g_{BL}$ is restricted at $10^{-5}$ level for $Q=0.1$. 
In both scenarios, $g_{BL}-M_{Z^\prime}$ is limited within a narrow region by the dark matter relic density, direct detection and baryon asymmetry constraints. 
\vspace{0.5cm}
\vspace{0.5cm}
\end{abstract}
\maketitle
\setcounter{footnote}{0}
\section{Introduction}
\label{sec:intro}

The observed baryon asymmetry of the universe (BAU) and dark matter (DM) have been long-standing problems of the particle physics. The BAU can be explained by a very attractive mechanism, the leptogenesis. This mechanism can arise naturally from the existence of the tiny neutrino masses. The heavy right-handed~(RH) neutrinos are added in the seesaw mechanisms to explain the light neutrino masses, while in the meantime satisfying the Sakharov conditions \cite{Sakharov:1967dj} to generate the lepton asymmetry, and transfer into the BAU via the sphaleron processes \cite{Klinkhamer:1984di}. If the masses of the right-handed neutrinos (RHNs) are introduced via a new $U(1)$ gauge, such as the $U(1)_{B-L}$, the seesaw can be explained naturally~\cite{Mohapatra:1980qe}. The $U(1)$ gauge also introduces an additional $Z^\prime$ boson. Once considered, it can lead to the scattering processes of RH neutrinos into SM fermions, therefore diluting the final BAU~\cite{Plumacher:1996kc}. If there exists $U(1)$ charged DM fermions $\chi$, the $Z^\prime$ boson can also contribute to the processes of DM-RHNs, so that make difference in 
the DM relic density and baryon asymmetry results.  In another word, the baryon asymmetry can  interplay the DM relic via the new $U(1)$ gauge. 

In such models, the DM production can be produced via the $Z^\prime$ mediated co-annihilation of the fermions and the RH neutrinos, $f\bar{f}$, $NN \to Z^\prime \to \chi \bar{\chi}$, while the similar couplings introduce the scattering process of $NN \to Z^\prime \to f\bar{f}$, $\chi\bar{\chi}$ during leptogenesis. As the DM relic density is more likely to be explained with stronger coupled $Z^\prime$ boson, while the observed BAU favors for weakly coupled ones, the scenarios where both the BAU and the DM relic density can be explained  simultaneously becomes non-trivial, and such conditions limits the choices for the parameter space. Therefore, it is interesting to look for such scenarios. 

Dark matter from leptogenesis has been discussed in Ref.~\cite{Falkowski:2011xh}, and followed by many other studies~\cite{Chun:2011cc,Davidson:2012fn, Falkowski:2017uya,Liu:2020mxj,Chang:2021ose,VanDong:2018yae,Narendra:2018vfw, Kuismanen:2012iz,Arina:2011cu,DAgnolo:2018wcn, Chianese:2018dsz,Chianese:2019epo,Biswas:2017tce,Qi:2022kgs,Qi:2022fzs}. Beside, the DM models with a $U(1)$ gauge is also studied at Ref.~\cite{Okada:2021nwo,Bandyopadhyay:2020ufc,Mohapatra:2020bze,Han:2020oet,Mohapatra:2019ysk,Heeba:2019jho,Abdallah:2019svm,delaVega:2021wpx,Seto:2020udg,Okada:2020cue,Escudero:2018fwn,Han:2018zcn,Okada:2018ktp,Cai:2018nob,Cox:2017rgn,Bandyopadhyay:2017bgh,Oda:2017kwl,Okada:2016tci,Biswas:2016bfo,Qi:2021rhh}.
Among them, Ref.~\cite{DAgnolo:2018wcn, Chianese:2018dsz, Chianese:2019epo,Biswas:2017tce} have considered DM production via the co-annihilation of the RH neutrinos, while the connection to the observed BAU is not yet discussed.

In this work, we discuss the interplay between dark matter, $Z^\prime$ and leptogenesis in the $B-L$ model. The paper is arranged as follows. In section \ref{sec:2}, we describe the framework of the model including dark matter and leptogenesis. In section \ref{sec:3}, we consider the Boltzmann equation involving dark matter, and give the evolution of the baryon asymmetry. In section \ref{sec:4}, we discuss the dark matter relic density in our model. In order to obtain the correct dark matter relic density, we consider the Freeze-out \cite{Chiu:1966kg} and Freeze-in \cite{Hall:2009bx} cases, respectively, which depend on the interaction strength as well as dark matter initial density at the early universe. What's more, dark matter direct detection experiments also give stringent constraints on the parameter space in the case of the Freeze-out scenario. We concretely discuss the interplay between dark matter, $Z^\prime$ and leptogenesis in section \ref{sec:5}, and we summary in the last part.
\section{Framework}
\label{sec:2}

In this work, we consider a simple extension of the SM within the $B-L$ framework, where a $Z'$ boson couples to three right-handed neutrinos $N_i$ with $i\in {e,\mu,\tau}$.    One singlet scalar $\Phi$  with $B-L$ charge $+2$ is added to provide Majorana masses to $N_i$. We also assume a fermion $\chi$ carrying $B-L$ charge $Q$ with $Q$ not equal to $\pm 1$ so that $\chi$ can be the dark matter candidate in the model, the relevant lagrangian can be given by,
\begin{eqnarray}\label{eq:L} \nonumber
\mathcal{L} \supset -Q g_{BL} Z^\prime_\mu \bar{\chi}\gamma^\mu\chi - \sum\limits_f g_{BL} q^\prime_f Z^\prime_\mu\bar{f}\gamma^\mu f \\
	- g_{BL} Z^\prime_\mu N_i \gamma^\mu N_i - m_{\chi}\bar{\chi}\chi-\mathcal{L}_N
\end{eqnarray}
%\begin{align}
%\mathcal{L} \supset  -Qg_{BL}   Z^\prime_\mu \bar{\chi}\gamma^\mu\chi - \sum\limits_f g_{BL} q^\prime_f Z^\prime_\mu\bar{f}\gamma^\mu f -g_{BL} Z^\prime_\mu N_i \gamma^\mu N_i-m_{\chi}\bar{\chi}\chi-\mathcal{L}_N
%\label{eq:L}
%\end{align}
where  $g_{BL}$ is the $U(1)_{B-L}$ gauge coupling constant,  $q'_f$ represents SM fermions $B-L$ charge with $q'_f=-1$ for leptons and $q'_f=1/3$ for quarks, $m_{\chi}$ is the $\chi$ mass and $\mathcal{L}_N$ is the term related with neutrino mass \cite{Heeck:2016oda},
\begin{align}
 \mathcal{L}_N=-(y\bar{L}HN_i+\lambda_N\bar{N_i^c}N_i\Phi+h.c.)
\end{align}
suppressing flavor indices, where $H$ is the SM Higgs and $L$ represents the SM leptons. $\lambda_N$ can be assumed as diagonal with positive entries without losing generality. We have $\Phi=(v_b+S)/\sqrt{2}$ in unitary gauge, and $v_b$ is the vacuum expectation value inducing the $Z'$ mass $M_{Z'}=2g_{BL}v_b$ and the Majarona mass matrix $\mathcal{M}_N=\sqrt{2}\lambda_Nv_b$. We can generate light neutrino masses via the Type-I seesaw mechanism in a standard way after electroweak symmetry breaking.
   
In this work, we discuss the interplay between $Z'$ boson, dark matter and leptogenesis. We consider the three right-handed neutrino masses are highly degenerate so that the observed baryon asymmetry can be generated by resonant leptogenesis \cite{Pilaftsis:2003gt,Okada:2012fs}. We use $N$ to represent three right-handed neutrinos for simplicity. The related parameters in our model are given as followed,
   \begin{eqnarray}
   M_{Z'},g_{BL},m_{\chi},Q,m_N,\tilde{m},\epsilon_{CP}
   \end{eqnarray}
 where $m_N$ represents right-handed mass, $\tilde{m}$ is the effective neutrino mass and we choose $\tilde{m}=10^{-4} eV$ as the benchmark value, and $\epsilon_{CP}$ is the CP asymmetry parameter. Since the BAU is produced via the resonant leptogenesis, right-handed neutrino mass may not necessarily be much heavy and $\epsilon_{CP}$ can be regarded as a free parameter in our model.The BAU obtains contributions from three right-handed neutrinos when we assume
the right-handed neutrinos are degenerate. If the decay widths of the three right-handed
neutrinos are comparable, then the generated BAU should be three times that mere one
right-handed neutrino decay. On the other hand, if the decay widths have a hierarchy,
the CP asymmetries will also do so and the generated BAU can be dominated by one of
the right-handed neutrinos, so that a one-flavor discussion will be sufficient, and we
consider such a scenario in this work. According to \cite{Blanchet:2009bu}, the allowed range for leptogenesis in the $M_{Z'}-m_N$ is constrained strictly in the case of $m_N<M_{Z'}/2$. We consider the decoupled-scalar limit $M_S \rightarrow \infty$ and $M_{Z'}>2m_N$ so that new extra processes contributing to leptogenesis are the s-channel interactions including $NN \rightarrow \chi\bar{\chi}$ and $NN\rightarrow f\bar{f}$ \cite{Heeck:2016oda}. On the other hand, dark matter production is generated by $Z'$-mediated processes of $f\bar{f}\rightarrow \chi\bar{\chi}$ and $NN\rightarrow \chi\bar{\chi}$. To obtain the observed relic density,  we consider the Freeze-in case and Freeze-out case separately in this work. This depends on the strength of the interaction  and  whether the initial density of dark matter in the early universe  is negligible or not.
\section{Boltzmann equations and leptogenesis}
\label{sec:3}

In this section, we give the Boltzmann equations of $N$ as well as dark matter. The dark matter production is determined by $f\bar{f}\rightarrow \chi \bar{\chi}$ and $NN\rightarrow \chi\bar{\chi}$, where the later process can also connect with leptogenesis. The total cross section of dark matter $\sigma_{\chi}(s)=\sigma_{f\chi}(s)+\sigma_{N\chi}(s)$ can be given by,
 \begin{eqnarray}
\sigma_{f\chi}(s)&=& \frac{8(Qg_{BL}^2)^2}{12\pi}\frac{\sqrt{1-4m_{\chi}^2/s}(s+2m_{\chi}^2)}{(s-M_{Z'}^2)^2+M_{Z'}^2\Gamma_{Z'}^2} \\
\sigma_{N\chi}(s)&=&\frac{(Qg_{BL}^2)^2}{6\pi}\frac{\sqrt{s(s-4m_N^2)}}{(s-M_{Z'}^2)^2+ M_{Z'}^2\Gamma_{Z'}^2}
 \end{eqnarray}
where $\sigma_{f\chi}(s)$ is the cross section of $f\bar{f}\rightarrow \chi\bar{\chi}$, $\sigma_{N\chi}(s)$ represents the cross section of $NN\rightarrow \chi\bar{\chi}$ and $s$ is the squared center of mass energy. The $Z'$ decay width $\Gamma_{Z'}$ is given by,
\begin{eqnarray}\nonumber
	\Gamma_{Z'}&=&\frac{g_{BL}^2 M_{Z'}}{24 \pi }(13+3(1-\frac{4{m_N}^2}{{M_{Z'}}^2})^{\frac{3}{2}}) \\
	     &+&\frac{Q^2g_{BL}^2M_{Z'}}{8\pi}(1-\frac{4m_{\chi}^2}{M_{Z'}^2})^{\frac{3}{2}}
\end{eqnarray}
Correspondingly, the reaction rate $\gamma_{f\chi}(z_1)$ and $\gamma_{N\chi}(z)$ are,
\begin{eqnarray} \nonumber
\gamma_{f\chi}(z_1)&=&\frac{m_N}{64\pi^4 z_1} \times \\ 
	&&\int_{4m_{\chi}^2}^{\infty}2(s-4m_{\chi}^2)\sigma_{f\chi}(s)\sqrt{s}K_1(\sqrt{s}\frac{z_1}{m_N})ds ~~~~ \\
\gamma_{N\chi}(z)&=&\frac{m_N^4}{64\pi^4 z}\int_{4}^{\infty}\hat{\sigma_{N\chi}}(y)\sqrt{y}K_1(z\sqrt{y})dy
\end{eqnarray}
where $K_1(z)$ is the modified Bessel function, $y$ is defined by $y=s/m_N^2$, $z_1=m_{\chi}/T$ and $z=m_N/T$ with $T$ being the  temperature. In addition, the reduced cross section $\hat{\sigma_{N\chi}}(y)$ is given by,
  \begin{eqnarray}
  \hat{\sigma_{N\chi}}(y)=\frac{(Qg_{BL}^2)^2}{6\pi}\frac{m_N^4\sqrt{(y-4)^3y}}{(M_{Z'}^2-m_N^2y)^2+M_{Z'}^2\Gamma_{Z'}^2}
  \end{eqnarray}
The Boltzmann equations to describe $N$ abundance $Y_N$, dark matter abundance $Y_X$ and $(B-L)$ asymmetry $Y_{B-L}$ are given as followed,
  \begin{eqnarray}\nonumber
	  && \frac{s_NH_N}{z^4}Y'_N = -(\frac{Y_N}{Y_{Neq}}-1)(\gamma_D +2\gamma_{hs}+4\gamma_{ht}) \\
	  && ~~~~~ -(\frac{Y_N^2}{Y_{Neq}^2}-\frac{Y_X^2}{Y_{Xeq}^2})2\gamma_{N\chi} 
  -(\frac{Y_N^2}{Y_{Neq}^2}-1)2\gamma_{NN} \\ \nonumber
	  && \frac{s_NH_N}{z^4}Y'_{B-L} = -(\frac{Y_{B-L}}{2Y_{Leq}}+\epsilon_{CP}(\frac{Y_N}{Y_{Neq}}-1))\gamma_D \\
	  && ~~~~~ -\frac{Y_{B-L}}{Y_{Leq}}(2(\gamma_N +\gamma_{Nt}+\gamma_{ht})+\frac{Y_N}{Y_{Neq}}\gamma_{hs}) \\
	  && \frac{s_NH_N}{z^4}Y'_X = -(\frac{Y_X^2}{Y_{Xeq}^2}-\frac{Y_N^2}{Y^2_{Neq}})2\gamma_{N\chi}  -(\frac{Y_X^2}{Y_{Xeq}^2}-1)\gamma_{f\chi} ~~~~~~
  \end{eqnarray}
About the quantities in the Boltzmann equations, we have,
 \begin{eqnarray}
	 s_N &=& \frac{2\pi^2}{45}g_{*}m_N^3, \\
	 H_N &=&\frac{m_N^2}{2\xi m_{\phi}}
 \end{eqnarray}
where $m_{pl}$ is the Planck mass with $m_{pl}=1,22\times10^{19}$ GeV, $g_{*}=106.75$ is the effective degrees of freedom and $\xi$ is defined by,
 \begin{eqnarray}
 \xi=\frac{1}{4\pi}\sqrt{\frac{45}{\pi g_{*}}}
 \end{eqnarray}
 $Y_{Neq}$ and $Y_{Xeq}$ describe the abundance of $N$ and dark matter at thermal equilibrium \cite{Davidson:2008bu},
 \begin{eqnarray}
	 Y_{Neq}(z) &=& \frac{45z^2}{2\pi^4 g_*}K_2(z), \\
	 Y_{Xeq}(z) &=& \frac{45z^2m_{\chi^2}}{m_N^22\pi^4 g_*}K_2(z\frac{m_{\chi}}{m_N})
 \end{eqnarray}
 $Y_{Leq}$ is the lepton abundance at thermal equilibrium with
 \begin{eqnarray}
 Y_{Leq}=\frac{6}{s_N}\frac{m_N^3\zeta(3)}{4\pi^2}
 \end{eqnarray}
 where $\zeta(x)$ is the Riemann zeta function. $\gamma_{NN}$ is the reaction rate of $NN\rightarrow Z' \rightarrow f\bar{f}$ and can be given by,
 \begin{eqnarray}
 \gamma_{NN}(z)=\frac{m_N^4}{64\pi^4 z}\int_4^{\infty}\hat{\sigma_{NN}}(y)\sqrt{y}K_1(z\sqrt{y})
 \end{eqnarray}
where the reduced cross section $\hat{\sigma_{NN}}$ is defined by,
 \begin{eqnarray}
 \hat{\sigma_{NN}}(y)=\frac{13g_{BL}^4}{6\pi}\frac{m_N^4\sqrt{(y-4)^3y}}{(M_{Z'}^2-m_N^2y)^2+M_{Z'}^2\Gamma_{Z'}^2}
 \end{eqnarray} 
 As for other terms in the Boltzmann equations, we give the explicit expression in the appendix \ref{sec:app}. The generated $(B-L)$ asymmetry can be converted into a baryon asymmetry by the sphelaron processes \cite{Khlebnikov:1988sr}, which is given by $Y_B=\frac{28}{79}Y_{B-L}$.
 
\begin{figure*}[htpb]
\centering
\begin{minipage}[t]{0.48\textwidth}
\centering
\includegraphics[width=7cm,height=5cm]{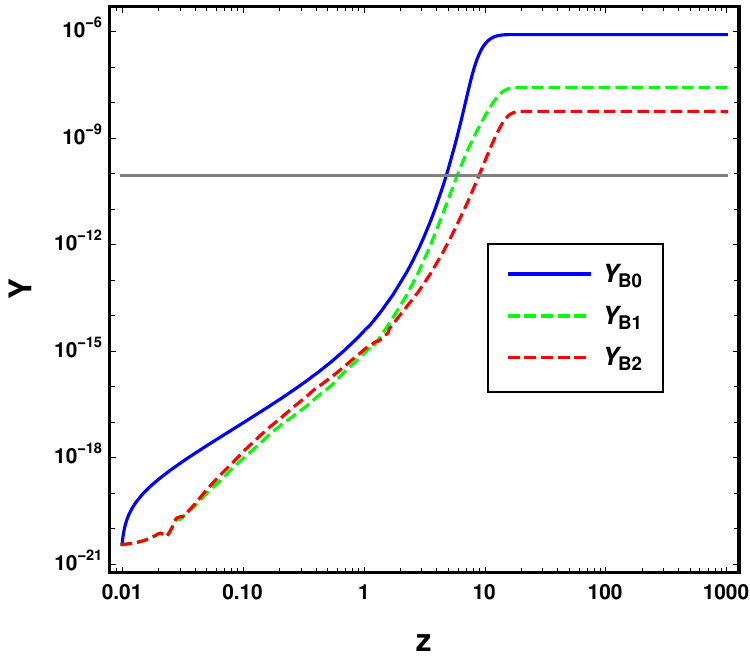}
\caption{Results of baryon asymmetry in the model, where we fix $Q=100$, $g_{BL}=0.01$, $|\epsilon_{CP}|=0.1$, $m_{\chi}=300$ GeV and $m_N= 1$ TeV. We take $M_{Z'}=4$ TeV, 4 TeV and 3 TeV corresponding to the blue, green and red lines respectively. Especially, the blue line is the result without dark matter. The grey line represents the observed baryon symmetry value $Y_B \approx 8.7 \times 10^{-11}$.}\label{Fig:fig1}
\end{minipage}
\begin{minipage}[t]{0.48\textwidth}
\centering
\includegraphics[width=7cm,height=5cm]{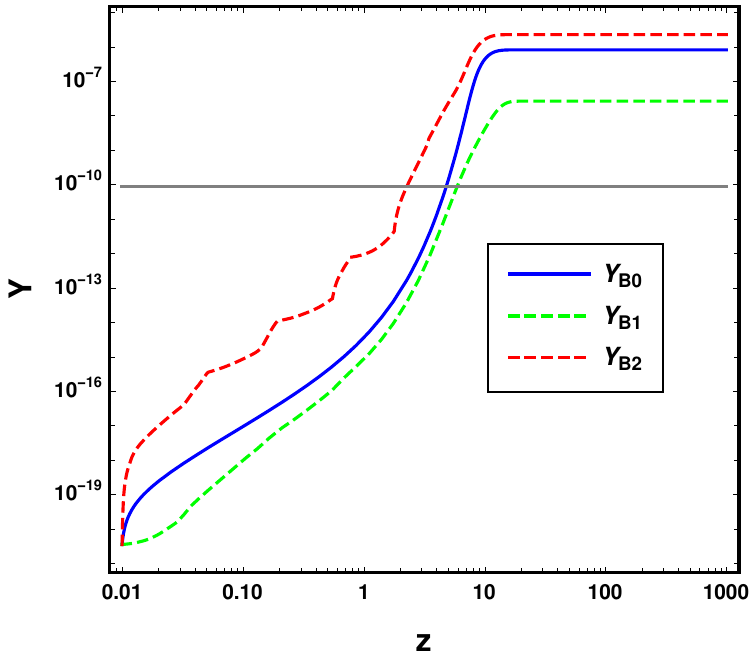}
\caption{Results of baryon asymmetry in the model, where we fix $Q=100$, $M_{Z'}=4$ TeV, $|\epsilon_{CP}|=0.1$, $m_{\chi}=300$ GeV and $m_N= 1$ TeV. We take $g_{BL}=0.01$, 0.01 and 0.001 corresponding to the blue, green and red lines respectively. Especially, the blue line is the result without dark matter. The grey line represents the observed baryon symmetry value $Y_B \approx 8.7 \times 10^{-11}$.}
\label{Fig:fig2}
\end{minipage}
\end{figure*}

  We give the results  in Fig.~\ref{Fig:fig1} and  Fig.~\ref{Fig:fig2}, where the grey lines in the two pictures are the observed baryon asymmetry value $Y_B \approx 8.7 \times 10^{-11}$  \cite{Planck:2018vyg} and the blue lines correspond to the case without dark matter. In both pictures, we set $Q=100$, $|\epsilon_{CP}|=0.1$, $m_{\chi}=300$ GeV and $m_N= 1$ TeV. In Fig.~\ref{Fig:fig1}, we fix $g_{BL}=0.01$ and take $M_{Z'}=4$ TeV, 4 TeV and 3 TeV corresponding to the blue, green and red lines respectively. In Fig.~\ref{Fig:fig2}, we fix $M_{Z'}=4$ TeV and vary $g_{BL}=0.01$, 0.01 and 0.001 corresponding to the blue, green and red lines respectively. According to Fig.~\ref{Fig:fig1}, the new process $NN \rightarrow \chi\bar{\chi}$ will further dilute the baryon  asymmetry besides other $Z'$-mediated processes related with $N$, and a lighter $M_{Z'}$ can correspond to a large cross section so that the dilution effect will be more efficient. Similarly, the coupling $g_{BL}$ can also determine these processes, and a larger $g_{BL}$ can also promote the dilution effect and give a smaller baryon asymmetry as we can see from Fig.~\ref{Fig:fig2}. Therefore, dark matter, $Z'$ and leptogenesis can be interplayed by the observed dark matter relic density and BAU in the model. We note for a much small $g_{BL}$ corresponding to the FIMP case, the contribution of dark matter to baryon asymmetry can be  negligible. However, dark matter cross section also depends on $g_{BL}$ and $M_{Z'}$ so that dark matter and leptogenesis can be connected by $Z'$. 

\section{Dark matter relic density}
\label{sec:4}

The evolution of Boltzmann equations can describe dark matter abundance and current
experiment analysis gives dark matter relic density with \cite{Planck:2018vyg},
 \begin{eqnarray}
 \Omega h^2=0.120\pm 0.001
 \end{eqnarray}
The dark matter density can be estimated by
   \cite{Backovic:2013dpa},
 \begin{eqnarray}
 \Omega_{\chi}h^2=\frac{m_{\chi}Y_{X,\infty}s_0}{\rho_{c,0}}\label{eqom}
 \end{eqnarray}
  where $Y_{X,\infty}$ is dark matter abundance at $z\rightarrow \infty$, $ s_0 \approx \rm 2889.2cm^{-3}$ is the present-day entropy density, and $ \rho_{c,0}/h^2=\rm 1.05 \times 10^{-5} GeV^2cm^{-3}$ is the critical density. The relic density can be obtained via the Freeze-in mechanism and Freeze-out mechanism, depending on the interaction strength as well as the initial density of dark matter at the early universe. Therefore, we divide two parts to discuss dark matter.

\subsection{Freeze-in dark matter}
\begin{figure}[h]
\centering
\begin{minipage}[t]{0.48\textwidth}
\centering
\includegraphics[width=7cm,height=5cm]{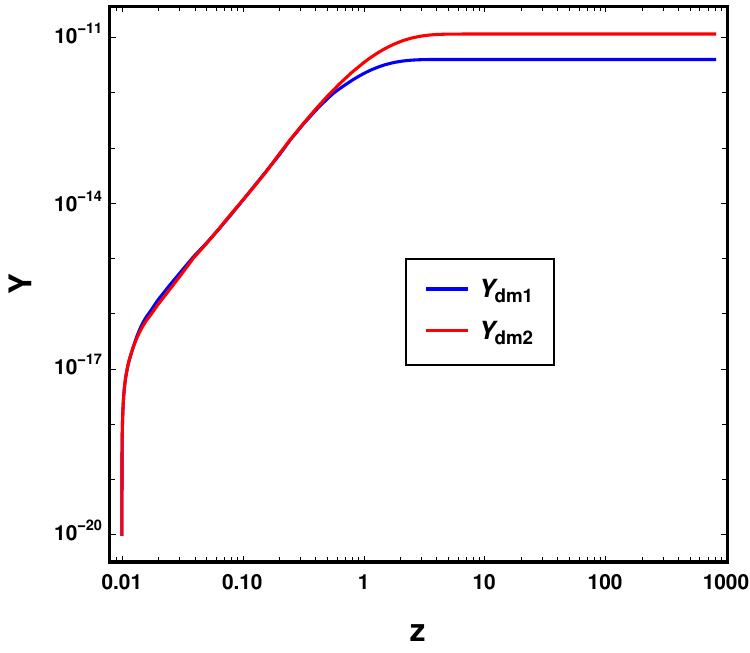}
\caption{Evolution of dark matter abundance where we fix $m_{\chi}=300$ GeV,$g_{BL}=10^{-8}$, $Q=100$, $|\epsilon_{CP}|=1$, $m_N=1$ TeV. The blue line corresponds to the case of $M_{Z'}=4$ TeV while the red line is $M_{Z'}=3$ TeV.}
\label{Fig:fig3}
\end{minipage}
\begin{minipage}[t]{0.48\textwidth}
\centering
\includegraphics[width=7cm,height=5cm]{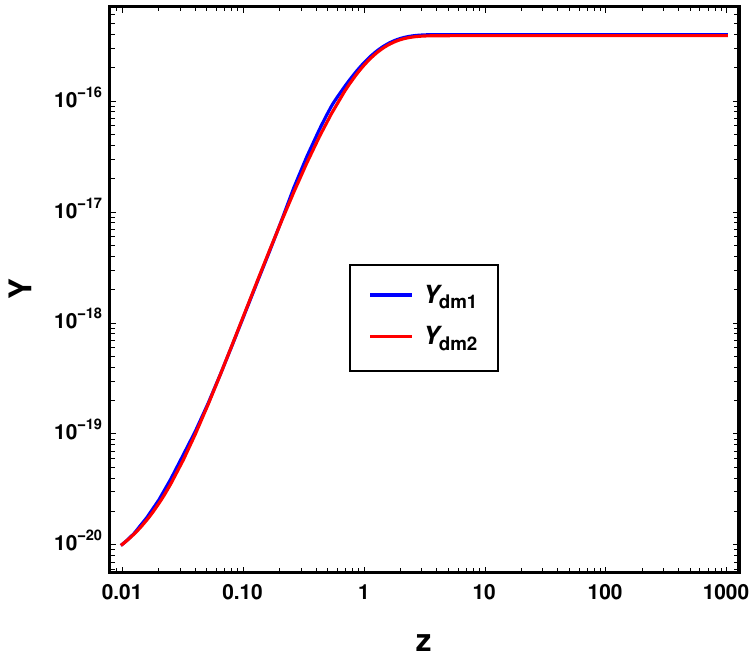}
\caption{Evolution of dark matter abundance where we fix $M_{Z'}=4$ TeV, $Q=100$, $|\epsilon_{CP}|=1$, $m_N=1$ TeV and $g_{BL}=10^{-8}$. The blue line corresponds to the case of $m_{\chi}=300$ GeV while the red line is $m_{\chi}=500$ GeV. }
\label{Fig:fig4}
\end{minipage}
\end{figure}

 In this part, we discuss the Freeze-in case in the model, this requires a small interaction coupling $Qg_{BL}$ and the initial dark matter density negligible. The results of dark matter abundance are given as followed. In Fig.~\ref{Fig:fig3}, we set $m_{\chi}=300$ GeV, $g_{BL}=10^{-8}$, $Q=100$, $|\epsilon_{CP}|=1$, $m_N=1$ TeV. The blue line corresponds to the case of $M_{Z'}=4$ TeV while the red line is $M_{Z'}=3$ TeV. The dark matter abundance will be larger with the interaction strength increases in the case of the Freeze-in scenario, and for a lighter $M_{Z'}$, we will have a larger cross section so that more dark matter. In Fig.~\ref{Fig:fig4}, we fix $M_{Z'}=4$ TeV, $Q=100$,   $|\epsilon_{CP}|=1$, $m_N=1$ TeV and $g_{BL}=10^{-8}$. The blue line corresponds to the case of $m_{\chi}=300$ GeV and the red line is $m_{\chi}=500$ GeV.  We will have different relic density according to Eq.~\ref{eqom} though the two lines almost coincide with each other.

\subsection{Freeze-out dark matter}
\begin{figure}[h]
\centering
\begin{minipage}[t]{0.48\textwidth}
\centering
\includegraphics[width=7cm,height=5cm]{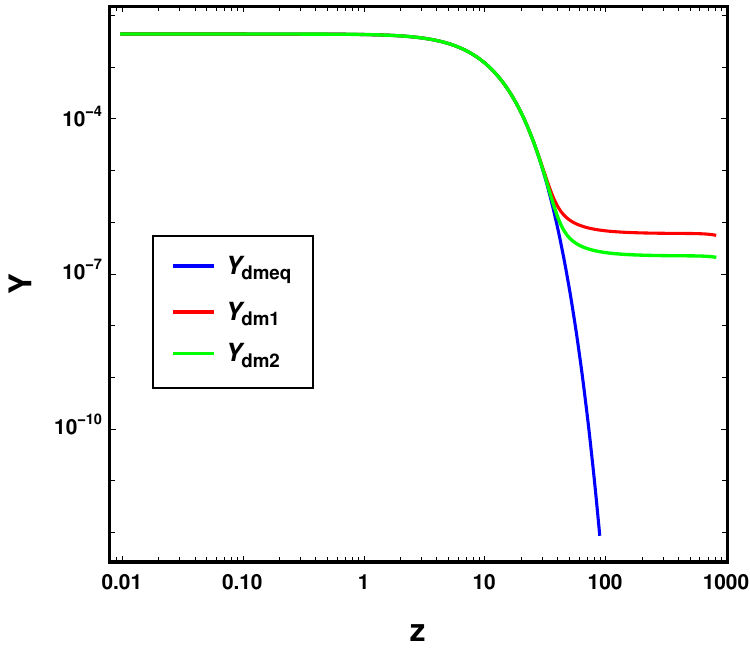}
\caption{ Evolution of dark matter abundance where we set $m_{\chi}=300 $ GeV, $Q=100$, $g_{BL}=0.01$, $m_N=1$ TeV and $|\epsilon_{CP}|=1$. The blue line is the evolution of dark matter abundance at thermally equilibrium, while the red and green lines correspond to $M_{Z'}=4$ TeV and 3 TeV respectively.}
\label{Fig:fig5}
\end{minipage}
\begin{minipage}[t]{0.48\textwidth}
\centering
\includegraphics[width=7cm,height=5cm]{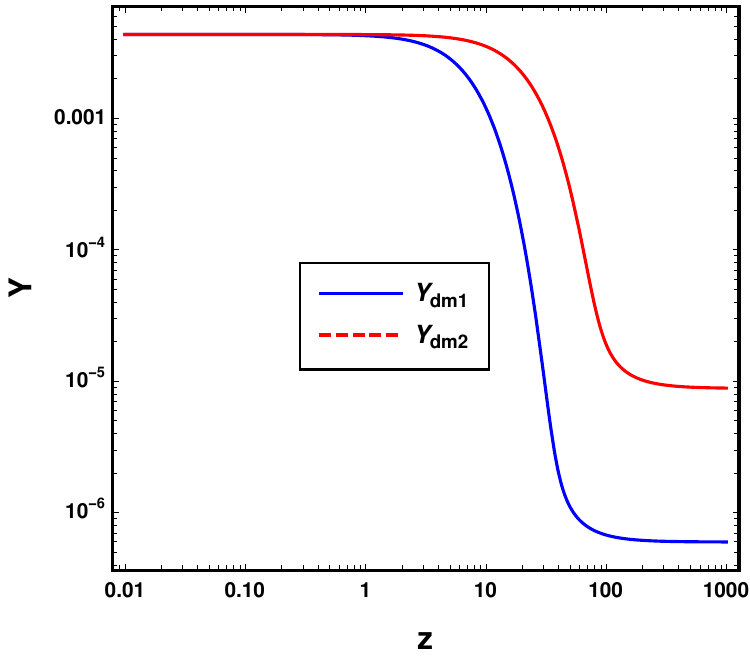}
\caption{Evolution of dark matter abundance where we set $Q=100$, $g_{BL}=0.01$ ,$M_{Z'}=4$ TeV, $m_N=1$ TeV and $|\epsilon_{CP}|=1$. The blue line corresponds to the case of $m_{\chi}=300$ GeV and the red is $m_{\chi}=100$ GeV.}
\label{Fig:fig6}
\end{minipage}
\end{figure}

 In this part, we discuss the Freeze-out scenario, where dark matter reached thermal equilibrium at the early universe and freeze out later. In Fig.~\ref{Fig:fig5}, we set $m_{\chi}=300$ GeV, $g_{BL}=0.01$, $Q=100$, $|\epsilon_{CP}|=1$, $m_N=1$ TeV. The blue line is the evolution of dark matter abundance at thermal equilibrium, while the red and green lines correspond to $M_{Z'}=4$ TeV and 3 TeV respectively. The freeze out occurs at about $z \approx 25$ and we obtain the relic density later. Contrary to the Freeze-in case, dark matter abundance will decrease when the interaction strength increase. In our model, a lighter $M_{Z'}$ always indicates a larger cross section of dark matter so that we have less abundance as we can see from Fig.~\ref{Fig:fig5}. In Fig.~\ref{Fig:fig6}, we set $Q=100$, $g_{BL}=0.01$, $M_{Z'}=4$ TeV, $m_N=1$ TeV and $|\epsilon_{CP}|=1$. The blue line corresponds to the case of $m_{\chi}=300$ GeV and the red is $m_{\chi}=500$ GeV. Similarly, we have less dark matter abundance in the case of a heavier dark matter mass, which induces a larger cross section.
 As for the Freeze-out case, the allowed parameter space is also constrained by direct detection for dark matter\cite{Nath:2021uqb}. Concretely speaking, for dark matter mass $m_{\chi} \gtrsim 2$ GeV, the direct detection experiments give the most stringent constraint on the spin-independent DM matter scattering with nucleon. Dark side-50 \cite{DarkSide:2018bpj}, PandaX-II \cite{PandaX-II:2017hlx} and LUX \cite{LUX:2019npm} experiments give the most stringent bounds in the case of 2 GeV $ \lesssim m_{\chi} \lesssim 6$ GeV, while XENON1T \cite{XENON:2020gfr} gives a stringent bound for $m_{\chi} >$ 6 GeV. On the other hand, since there is no evidence for the existence of WIMP DM taking current searches into consideration, it is possible that the  WIMP-nucleus scattering cross section bound is at $10^{-13}$ pb level in the case of dark matter mass larger than $100$ GeV with the neutrino-floor limit according to \cite{Billard:2013qya,Billard:2021uyg}, which can put the most stringent constraint on the dark matter parameter space. In this work, the scattering is arising from the exchange of $Z'$ between dark matter and nucleon and the DM-nucleon cross section $\sigma_{SI}$ can be estimated by \cite{Nath:2021uqb},
 \begin{eqnarray}
 \sigma_{SI} \approx \frac{1}{\pi}Q^2g_{BL}^4\frac{\mu^2}{M_{Z'}^4}
 \end{eqnarray}
  where $\mu=m_{\chi}m_{Nu}/(m_{\chi}+m_{Nu})$ is the reduced mass for the DM-nucleon with $m_{Nu}=0.983$ GeV the nucleon mass.

\section{ Interplay between dark matter, leptogenesis and $Z^\prime$ }
\label{sec:5}

In this section, we discuss the interplay between dark matter, leptogenesis and $Z^\prime$, we divide three parts as follows.

\begin{figure}[h]
\centering
\begin{minipage}[t]{0.48\textwidth}
\centering
\includegraphics[width=7cm,height=5cm]{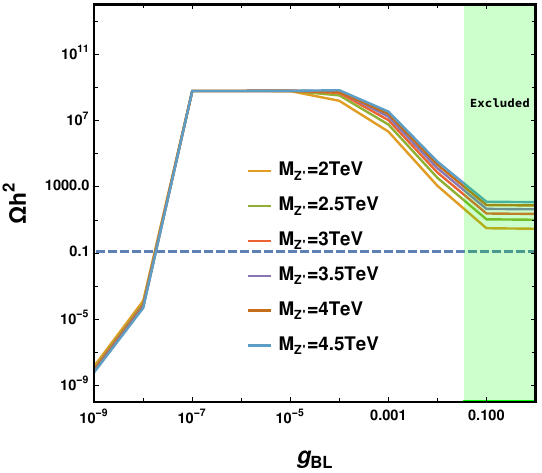}
\caption{Results of dark matter relic density, where the blue dashed line is the current observed relic density with $\Omega h^2=0.12$, the green region is excluded by unitary constraint. We fix $Q=100$, $m_N=1/4 M_{Z'}$ and $|\epsilon_{CP}|=1$, the colored lines correspond to $M_{Z'}$ ranging from 2 TeV to 4.5 TeV.}
\label{Fig:fig7}
\end{minipage}
\begin{minipage}[t]{0.48\textwidth}
\centering
\includegraphics[width=7cm,height=5cm]{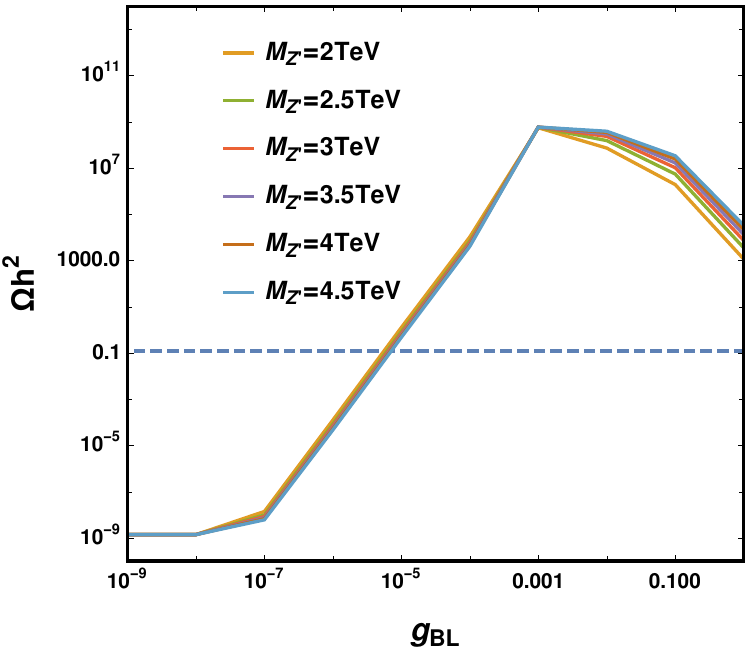}
\caption{Results of dark matter relic density, where the blue dashed line is the current observed relic density with $\Omega h^2=0.12$. We fix $Q=0.01$, $m_N=1/4 M_{Z'}$ and $|\epsilon_{CP}|=1$, the colored lines correspond to $M_{Z'}$ ranging from 2 TeV to 4.5 TeV.}
\label{Fig:fig8}
\end{minipage}
\end{figure}

\paragraph{Interplay between $Z^\prime$ and DM}
~\\ 
$M_{Z'}$ and $g_{BL}$ can affect dark matter relic density according to the expression of dark matter cross section, and the results are given in Fig.~\ref{Fig:fig7} and Fig.~\ref{Fig:fig8}. We fix $Q=100$, $m_N=1/4 M_{Z'}$ and $|\epsilon_{CP}|=1$ in Fig~\ref{Fig:fig7}, while in Fig.~\ref{Fig:fig8} we set $Q=0.01$. According to Fig.~\ref{Fig:fig7}, the dashed blue line corresponds to the observed relic density $\Omega h^2=0.12$. For other colored lined, we set $M_{Z'}$ ranging from 2 TeV to 4.5 TeV. The Green region is excluded by unitary constraint with $Qg_{BL}<\sqrt{4\pi}$ \cite{Duerr:2016tmh}. For a fixed $M_{Z'}$, dark matter relic density increases with $g_{BL}$ increases in the case of $g_{BL}<10^{-7}$. This region corresponds to the Freeze-in dark matter that larger $g_{BL}$ induces a larger cross section and more dark matter. For $g_{BL}>10^{-4}$, dark matter relic density is generated by Freeze-out, and will decrease with increasing of $g_{BL}$. On the other hand, for the fixed $g_{BL}$, a lighter $M_{Z'}$ can induce a larger cross section for dark matter, and we have less(more) relic density for Freeze-out(Freeze-in) scenario. A similar conclusion can be found in Fig.~\ref{Fig:fig8}, but we have a wider parameter  space for $g_{BL}$ since $Q=0.01$.
  
\begin{figure}[h]
\centering
\begin{minipage}[t]{0.48\textwidth}
\centering
\includegraphics[width=7cm,height=5cm]{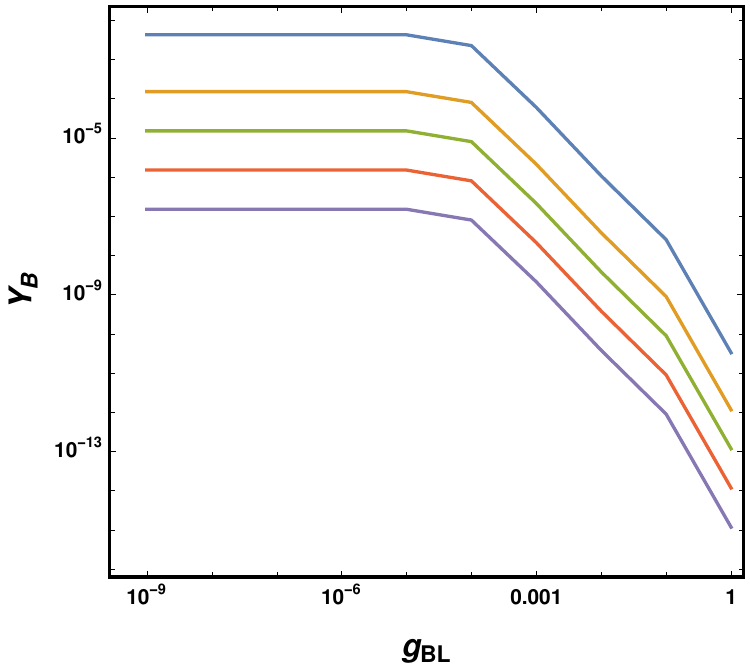}
\caption{ Results of baryon asymmetry related with $g_{BL}$, where we set $Q=0.01$, $M_{Z'}=4$ TeV, $m_N=1/3 M_{Z'}$ and $m_{\chi}=500$ GeV. The colored lines correspond to $|\epsilon_{CP}|$ equal to $1,0.1,0.01,0.001$ and 0.0001 respectively.}
\label{Fig:fig9}
\end{minipage}
\begin{minipage}[t]{0.48\textwidth}
\centering
\includegraphics[width=7cm,height=5cm]{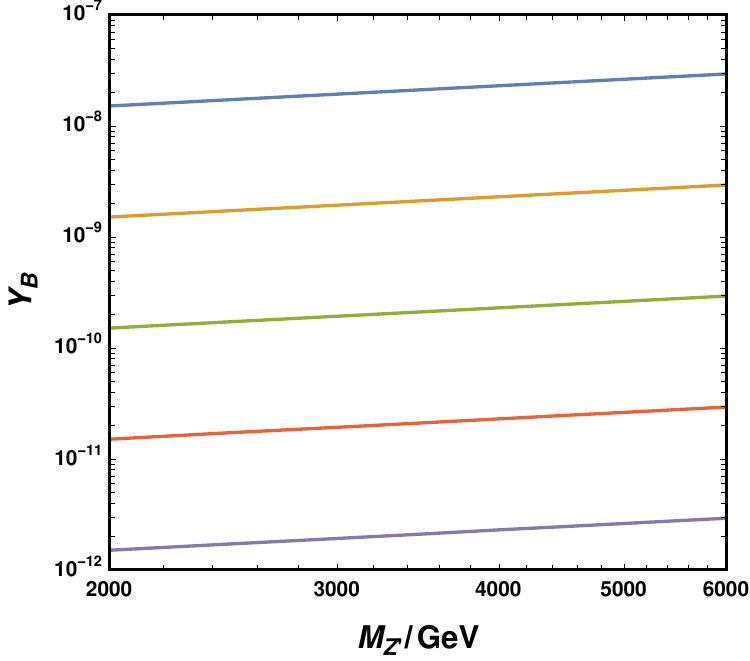}
\caption{Results of baryon asymmetry related with $M_{Z'}$, where we set $g_{BL}=0.01$, $m_{\chi}=100$ GeV, $Q=0.01$ and $m_N=0.4M_{Z'}$. The colored lines correspond to $|\epsilon_{CP}|$ equal to $1,0.1,0.01,0.001$ and 0.0001 respectively.}
\label{Fig:fig10}
\end{minipage}
\end{figure}

\paragraph{Interplay between $Z^\prime$ and leptogenesis}
~\\ 
The existence of $Z'$ introduces new processes related with $N$ as well as dark matter in the model, which can dilute the result of baryon asymmetry. In Fig.~\ref{Fig:fig9}, we set $Q=0.01$, $M_{Z'}=4$ TeV, $m_N=1/3 M_{Z'}$ and $m_{\chi}=500$ GeV. The colored lines correspond to $|\epsilon_{CP}|$ equal to $1,0.1,0.01,0.001$ and 0.0001 respectively. For the small $g_{BL}$ such as $g_{BL}<10^{-6}$. $Z'$-mediated processes can be negligible and make little difference on the baryon symmetry. However, with the increase of $g_{BL}$, when $g_{BL}>0.0001$, contribution of $Z'$-mediated processes are so larger that the dilution effect is  efficient and $Y_B$ decreases sharply. In addition, the contribution of $g_{BL}$ and $\epsilon_{CP}$ are opposite, and a proper baryon symmetry can be obtained either a pair of large  $(\epsilon_{CP},g_{BL})$ or small in the case of $g_{BL}>0.0001$. In Fig.~\ref{Fig:fig10}, we fix $g_{BL}=0.01$, $m_{\chi}=100$ GeV, $Q=0.01$ and $m_N=0.4M_{Z'}$. The colored lines correspond to $|\epsilon_{CP}|$ equal to $1,0.1,0.01,0.001$ and 0.0001 respectively. For a heavier $M_{Z'}$, the cross section related with $N$ is smaller so that we have a larger $Y_B$ as we can see from Fig.~\ref{Fig:fig10}. 

\begin{figure}[h]
\centering
\begin{minipage}[t]{0.48\textwidth}
\centering
\includegraphics[width=7cm,height=5cm]{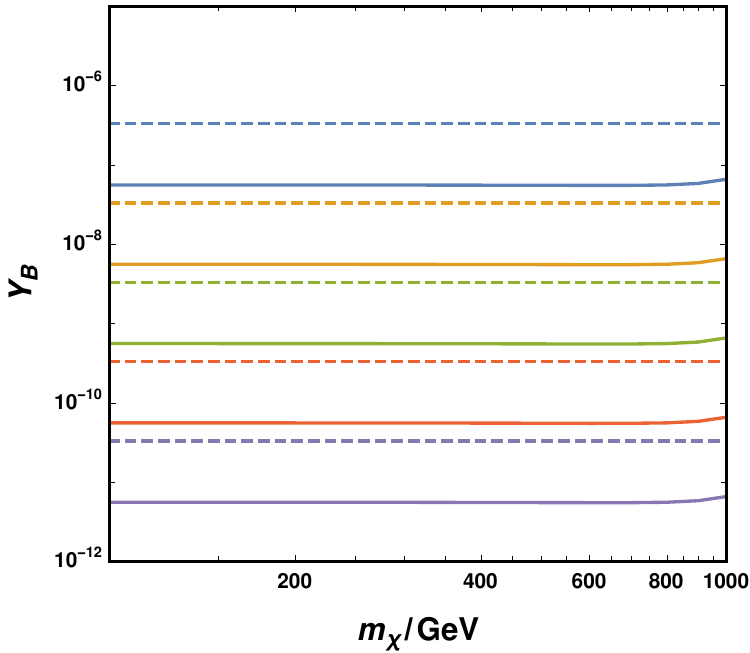}
\caption{ Evolution of baryon asymmetry related with $m_{\chi}$, where we set $Q=100$, $g_{BL}=0.01$, $M_{Z'}=3$ TeV and $m_N=1/3M_{Z'}$. The colored lines correspond to $|\epsilon_{CP}|$ equal to $1,0.1,0.01,0.001$ and 0.0001 respectively, and the same color but dashed lines represent the case without dark matter.}
\label{Fig:fig11}
\end{minipage}
\begin{minipage}[t]{0.48\textwidth}
\centering
\includegraphics[width=7cm,height=5cm]{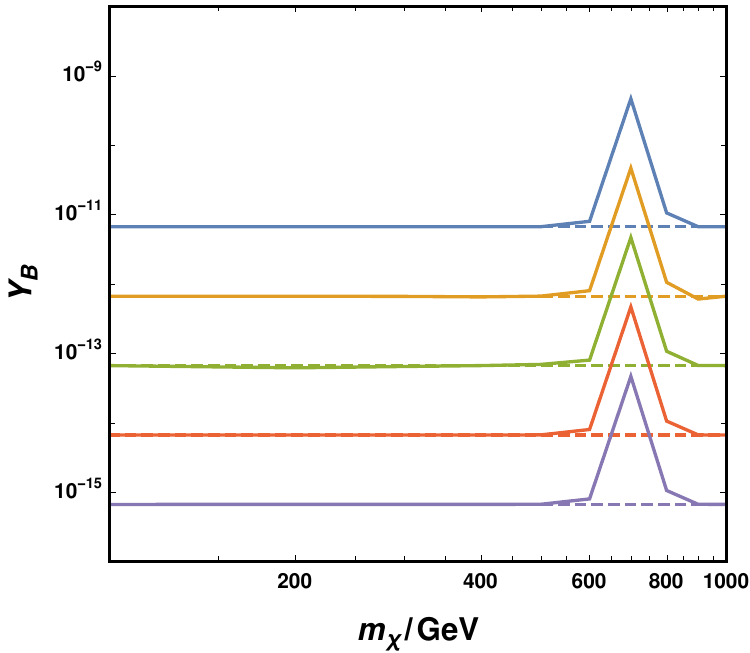}
\caption{Evolution of baryon asymmetry related with $m_{\chi}$, where we set $Q=100$, $g_{BL}=0.01$, $M_{Z'}=2$ TeV and $m_N=1/3M_{Z'}$. The colored lines correspond to $|\epsilon_{CP}|$ equal to $1,0.1,0.01,0.001$ and 0.0001 respectively, and the same color but dashed lines represent the case without dark matter.}
\label{Fig:fig12}
\end{minipage}
\end{figure}

\paragraph{Interplay between DM and leptogenesis}~\\ 
DM in our model is related to the s-channel processes  $f\bar{f}\rightarrow \chi \bar{\chi}$ and  $NN\rightarrow \chi\bar{\chi}$, and the later process can affect the results of baryon asymmetry. We note that the contribution of dark matter to baryon asymmetry can be negligible in the Freeze-in scenario due to the weak interaction strength and we focus on the case of Freeze-out.  In Fig.~\ref{Fig:fig11}, we consider the evolution of baryon asymmetry related to dark matter mass $m_{\chi}$. We set $Q=100$, $g_{BL}=0.01$, $M_{Z'}=3$ TeV and $m_N=1/3M_{Z'}$. The colored lines correspond to $|\epsilon_{CP}|$ equal to $1,0.1,0.01,0.001$ and 0.0001 respectively, and the same color but dashed lines represent the case without dark matter. According to Fig.~\ref{Fig:fig11},  baryon asymmetry is diluted efficiently due to  dark matter production. Therefore, one needs a larger $\epsilon_{CP}$ to obtain the correct baryon asymmetry compared with the case without dark matter. In Fig.~\ref{Fig:fig12}, we consider a lighter $M_{Z'}$ but other parameters are the same as in Fig.~\ref{Fig:fig12}. As we mentioned above, a lighter $M_{Z'}$ always induces a large interaction cross section related with $N$, and we have a lower $Y_B$ compared with Fig.~\ref{Fig:fig11}. However, the contribution of dark matter to baryon asymmetry is less efficient, and the same color lines almost coincide  with each other. An interesting point happens when $m_{\chi} \approx m_N$ and we found a peak in $Y_B$. Such a peak arises from the fact that leptogenesis and dark matter Freeze-out occurs at nearly the same time when $m_{\chi} \approx m_N$. More $N$ has been generated by the inverse process of $NN \rightarrow \chi \bar{\chi}$ so that we have a much larger $Y_B$. If $m_{\chi}\gg m_N$, freeze-out of dark matter  will be earlier than leptogenesis that $N$ can still be thermally equilibrium during dark matter freeze-out, and dark matter makes little difference on leptogenesis. For $m_{\chi}<m_N$, dark matter can still be thermally equilibrium during leptogenesis, the process related with dark matter will just keep $N$ close to equilibrium and dilute the baryon asymmetry like the SM fermions which depend on the interaction strength. Therefore, dark matter mass can make little difference on leptogenesis in the model except for $m_{\chi} \approx m_N$. 

\paragraph{ IV). Combined results}

   We give the combined results in this part, and we divide our discussion into two cases with Freeze-out dark matter and Freeze-in dark matter. For simplicity, we fix $Q$ with $Q=0.1$ in both cases, and we scan the parameter space with
    \begin{align*}
   M_{Z'} \subset  [1 \ \mathrm{TeV},5\ \mathrm{TeV}],   
    m_N \subset  [0.5 \ \mathrm{TeV},1.5 \ \mathrm{TeV}],
    \end{align*}
    \begin{align}
   g_{BL} \subset [ 10^{-8} ,1],
    m_{\chi} \subset [1\ \mathrm{GeV}, 1000\ \mathrm{GeV}]
 \end{align}
 where in the Freeze-out case, $g_{BL}$ takes a value ranging from $(10^{-5},1]$ and for the Freeze-in case, we consider $10^{-8} \leq g_{BL}\leq10^{-5}$.
 
\begin{figure}[h]
\centering
\begin{minipage}[t]{0.48\textwidth}
\centering
\includegraphics[width=7cm,height=5cm]{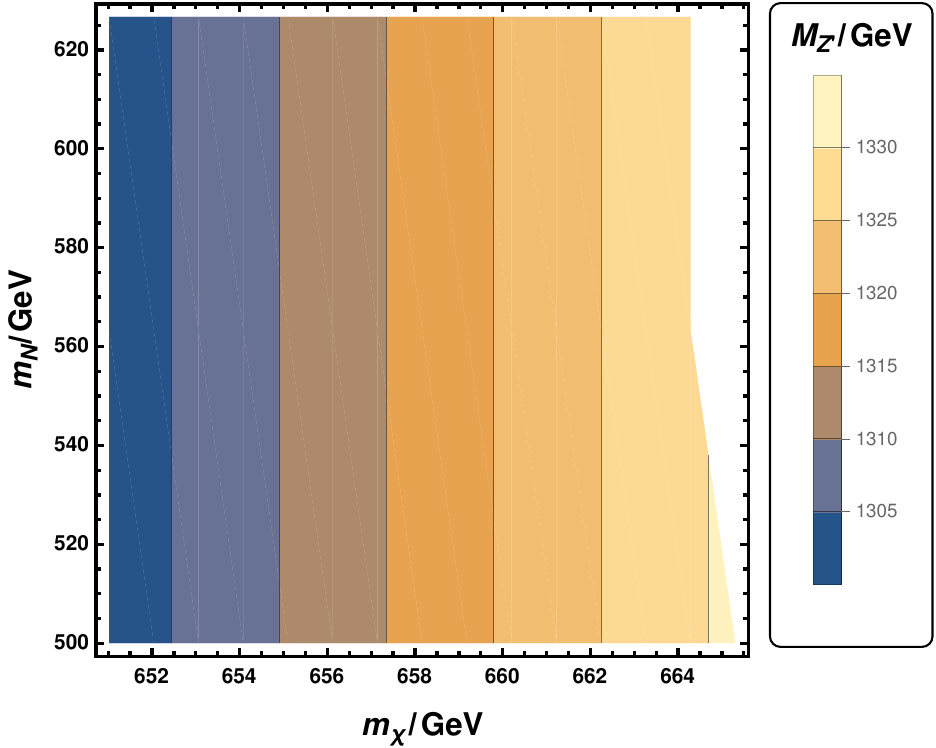}
\caption{Parameter space satisfying dark matter relic density as well as direct detection constraint.}
\label{Fig:fig13}
\end{minipage}
\begin{minipage}[t]{0.48\textwidth}
\centering
\includegraphics[width=7cm,height=5cm]{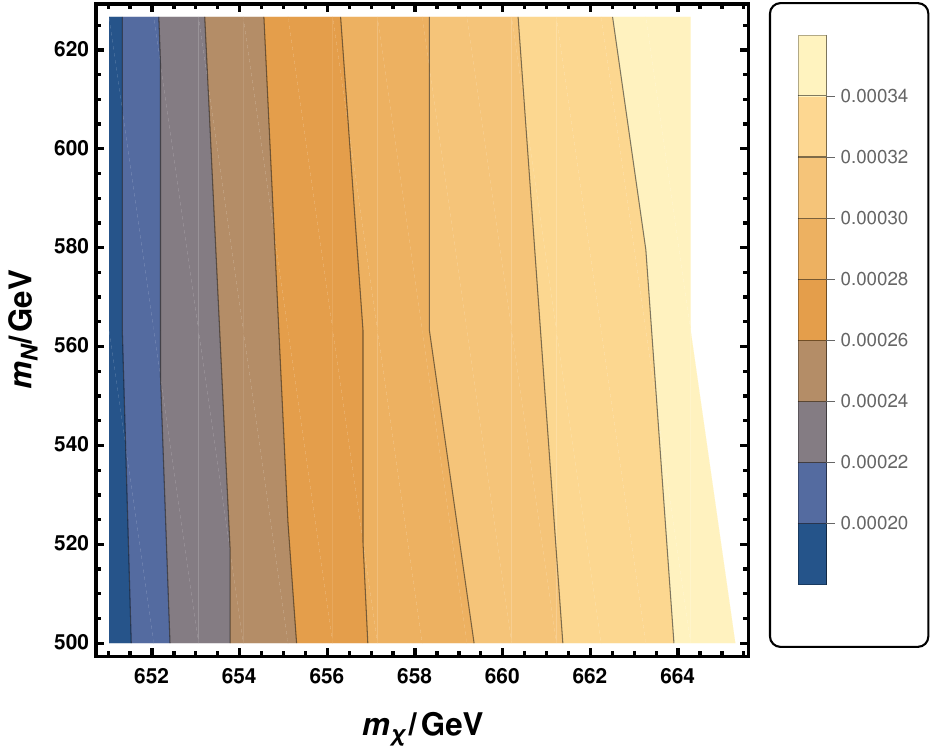}
\caption{Parameter space satisfying dark matter relic density as well as direct detection constraint, where the legends represent the viable value of $g_{BL}$.}
\label{Fig:fig14}
\end{minipage}
\end{figure}

\begin{figure}[h]
\centering
\begin{minipage}[t]{0.48\textwidth}
\centering
\includegraphics[width=7cm,height=5cm]{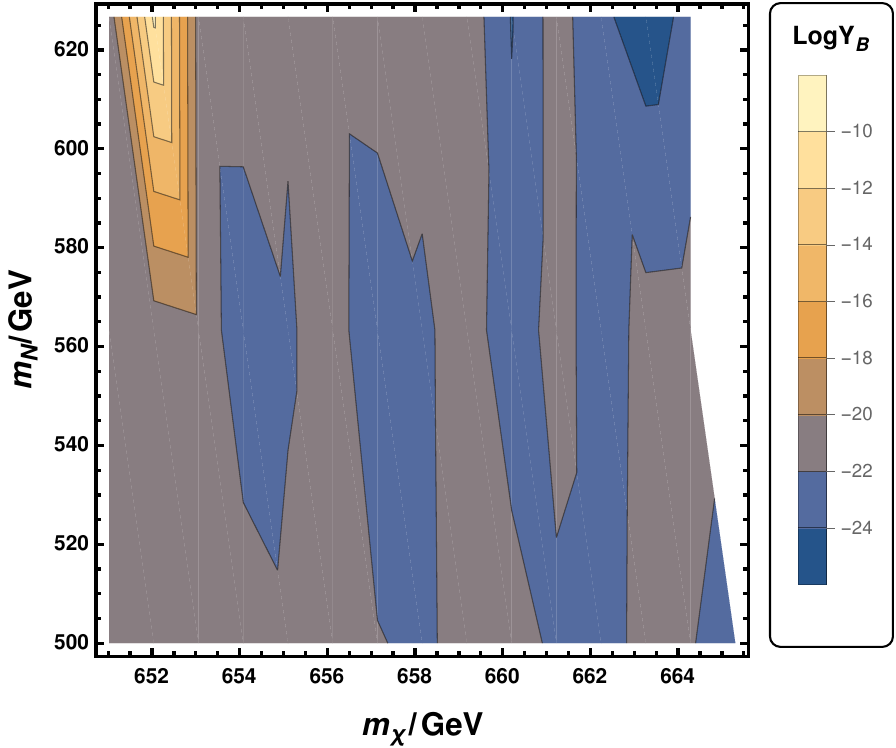}
\caption{Parameter space satisfying dark matter relic density as well as direct detection constraint,where the legends represent the viable value of $Log_{Y_B}$,and $\epsilon_{CP}$ is fixed with $\epsilon_{CP}=10^{-6}$.}
\label{Fig:fig15}
\end{minipage}
\begin{minipage}[t]{0.48\textwidth}
\centering
\includegraphics[width=7cm,height=5cm]{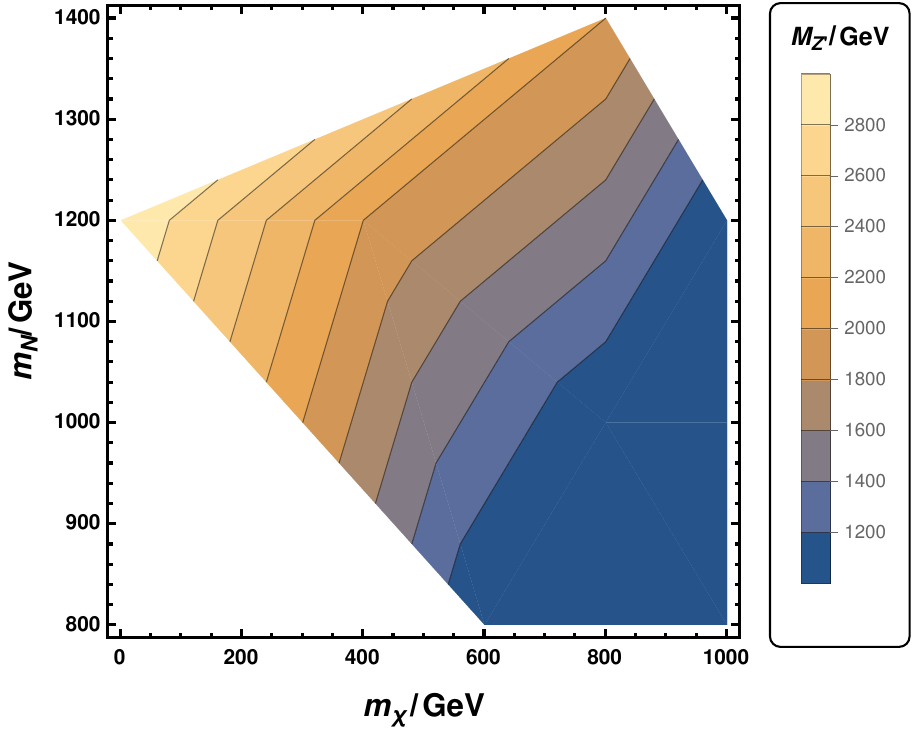}
\caption{Contour plot of $m_{\chi}-m_N$ with dark matter relic density in the case of Freeze-in, where $g_{BL}$ is fixed with $g_{BL}=10^{-5}$.}
\label{Fig:fig16}
\end{minipage}
\end{figure}

  In Fig.~\ref{Fig:fig13}, Fig.~\ref{Fig:fig14} and Fig.~\ref{Fig:fig15}, we give the results of the Freeze-out case. We use Micromegas \cite{Belanger:2013oya} to calculate the relic density numerically, and take the direct detection constraint into consideration with $\sigma_{SI}<10^{-13}$ pb in the case of $m_{\chi}>100$ GeV. According to Fig.~\ref{Fig:fig13} and Fig.~\ref{Fig:fig14},  the viable parameter space satisfying relic density constraint as well as direct detection constraint is $m_{\chi}\subset [650\ \rm GeV,666\ \rm GeV]$, $m_N \subset[500\ \rm GeV, 620\ \rm GeV]$, $M_{Z'} \subset [1300\ \rm GeV, 1330\ \rm GeV]$ and $g_{BL} \subset [2\times 10^{-4},3.4 \times 10^{-4}]$. Dark matter direct detection puts the most stringent constraint on the parameter space that the viable dark matter mass is $m_{\chi} \approx M_{Z'}/2$, which corresponds to the resonance-enhanced process of $NN \to Z' \to \chi\bar{\chi}$. In addition, $M_{Z'}$ is constrained within a narrow range at a TeV scale, for a lighter $Z'$, one can have a larger scattering cross section for DM-nucleon so that beyond the direct detection constraint. On the other hand, a heavier $Z'$ can correspond to a lower DM-nucleon cross section but lead to dark matter over-abundance.  The combined constraints on the $M_{Z'}-g_{BL}$  gives $M_{Z'}$ at $1.3$ TeV and $g_{BL}$ at $2\times 10^{-4}$ level, which also meets experiment data from ATLAS \cite{ATLAS:2014jlg} and LEP-II \cite{Buckley:2011vc,Carena:2004xs}. 
  
  In Fig.~\ref{Fig:fig15}, we fix $\epsilon_{CP}= 10^{-6}$ and give the contour plot of $m_{\chi}-m_N$ with $Log_{Y_B}$, which satisfies dark matter relic density as well as direct detection constraint. According to the astronaut experiments, the observed baryon asymmetry $Y_B$ is  $Y_B \approx 8 \times 10^{-11}$ \cite{Planck:2018vyg}, and $Log_{Y_B}$ value corresponds to about -23 in Fig.~\ref{Fig:fig15}.  For $m_{\chi}<653$ GeV and $m_N>570$ GeV, the baryon asymmetry $Y_B$ is smaller than the observed value. However, for $m_{\chi}>660$ GeV, the baryon asymmetry is strengthened due to the process of $\chi\bar{\chi}\to NN$, and one can obtain the correct baryon asymmetry as long as $m_N \subset[500$ GeV,620 GeV].
  
  As for the Freeze-in dark matter, we will have a simplified discussion compared with Freeze-out case. Since dark matter can make little difference on the result of baryon asymmetry due to the small interaction strength, we focus on the dark matter relic density and the result is given in Fig.~\ref{Fig:fig16}. In addition, dark matter production is suppressed by the small coupling $g_{BL}Q$ and heavy gauge boson mass $M_{Z'}$. The viable $g_{BL}$ to generate the correct relic density is at $10^{-5}$ level, while for smaller $g_{BL}$, dark matter relic density is much smaller than the observed value with the Freeze-in mechanism. According to Fig.~\ref{Fig:fig16}, we give the contour plot of $m_{\chi}-m_N$ with $M_{Z'}$ where we have fixed $g_{BL}=10^{-5}$. Unlike the freeze-out case, we have a wider parameter space for $m_{\chi}$, $m_N$ as well as $M_{Z'}$ with $m_{\chi} \subset [1$ GeV,1000 GeV], $m_N \subset [0.8$ TeV,1.4 TeV] and $M_{Z'} \subset [1.2$ TeV, 2.8 TeV]. What's more, with the increase of $m_{\chi}$, the viable parameter space for $m_N-M_{Z'}$ with correct relic density is more flexible. On the other hand, dark matter production arises from the processes $f\bar{f},NN \to Z' \to \chi\bar{\chi}$, and for the given $m_N$ value, a heavier $m_{\chi}$ always demands a lighter $Z'$ mass to obtain the right relic density.

\section{summary}
\label{sec:so}

  Dark matter, neutrino masses and baryon asymmetry are the long-standing SM unexplained problems, while the latter two can be explained by the so-called leptogenesis mechanism. By introducing right-handed neutrinos, tiny neutrino masses can be generated by a type-I seesaw mechanism, while CP-violation decays of out-of-equilibrium right-handed neutrinos can produce lepton asymmetry, 
and the lepton asymmetry can then be converted to baryon asymmetry during the sphelaron process.
   
   In this work, we consider a $B-L$ model to discuss  leptogenesis, dark matter and new gauge boson $Z'$. The baryon asymmetry is generated by resonant leptogenesis, where the right-handed neutrino masses are highly degenerate and a successful leptogenesis can be obtained at the TeV level. The fermion dark matter $\chi$ carries $U(1)_{B-L}$ charge but does not couple with other particles in the model, so dark matter production is mainly generated by the processes of $f\bar{f} \to Z' \to \chi\bar{\chi}$ and $NN \to Z' \to \chi\bar{\chi}$. To obtain the observed dark matter relic density, we consider the Freeze-out and Freeze-in mechanisms which correspond to two different cases. In the  case of Freeze-out, dark matter abundance will decrease with the increasing of interaction strength, while for the Freeze-in dark matter, abundance will increase when the interaction cross section increases. In addition, dark matter is also limited stringently by the direct detection experiment in the case of Freeze-out, and such constraints can be negligible for the Freeze-in dark matter due to the weak interaction strength.
   
    We consider the interplay between leptogenesis, dark matter and new gauge boson $Z'$ in the case of Freeze-out and Freeze-in, and we divide our discussion into three parts. Firstly, for the new gauge boson $Z'$ and dark matter, relic density is  determined by $M_{Z'}$ and $g_{BL}$.  For $Q=100$, we come to Freeze-in dark matter in the case of $g_{BL}<10^{-7}$ and Freeze-out dark matter for $g_{BL}>10^{-4}$. What's more, for a fixed $g_{BL}$, a lighter $M_{Z'}$ can induce a larger cross section for the dark matter so that we have less(more) relic density for Freeze-out (Freeze-in) scenario. Then, the existence of $Z'$ introduces new processes related with $N$ as well as dark matter in the model, which can dilute the result of baryon asymmetry. For the too small $g_{BL}$ and much heavy $M_{Z'}$, the cross section of $Z'$-mediated processes such as $NN \to Z' \to \chi\bar{\chi}$ is too small so that the dilution is obvious. However, with increasing of $g_{BL}$, such effect will be efficient and $Y_B$ decrease sharply in the case of $g_{BL}>0.0001$ for $M_{Z'}=4$ TeV. As for dark matter and leptogenesis, dark matter production in our model is related with the s-channel processes of $NN \to \chi \bar{\chi}$ and $f\bar{f} \to \chi \bar{\chi}$, and the contribution of dark matter to baryon asymmetry can be negligible in the Freeze-in scenario due to the weak interaction strength. We found dark matter makes little difference on leptogenesis in the model except $m_{\chi}\approx m_N$ when the baryon asymmetry is strengthened by the inverse process of $NN \to \chi \bar{\chi}$. We give the combined results by scanning over a viable parameter space, which satisfies the relic density constraint. For the freeze-out scenario, dark matter direct detection experiments give the most stringent limit on the parameter space, where dark matter is constrained with $m_{\chi}\approx M_{Z'}/2 $, $M_{Z'} \subset [1300$ GeV,1330 GeV], and $g_{BL} \subset [2 \times 10^{-4},3.4 \times 10^{-4}$], and we can obtain the right baryon asymmetry in the case of $m_{\chi} \subset [654$ GeV,664 GeV]. For the Freeze-in case, we focus on the dark matter constraint and we have a wider parameter space for dark matter mass and $M_{Z'}$ with $m_{\chi} \subset[1$ GeV,1000 GeV] and $M_{Z'} \subset [1.2$ TeV, 2.8 TeV], but the viable $g_{BL}$ parameter space is constrained within $10^{-5}$ level with $Q=0.1$, and for smaller $g_{BL}$, the relic density is much smaller than the observed value. Above all, in both Freeze-in and Freeze-out scenarios, one can find a viable parameter space satisfying dark matter relic density, direct detection and baryon asymmetry constraints. However, the parameter space is well limited and the coupling $g_{BL}$ and new gauge boson mass $M_{Z'}$ are constrained within a narrow region. Such regions are under current collider experiment constrain and can be tested in the future experiment on the search for heavy $Z'$ at LHC.

\appendix 

\section{Formulas  }
\label{sec:app}

In this part, we give other relevant expressions of the reaction rate in the Boltzmann equations \cite{Iso:2010mv}. 
Firstly, the reaction rate for $N$ decay $\gamma_{D}$ is given by,
\begin{eqnarray}
  \gamma_D = \frac{m_N^3}{\pi^2 z}K_1(z)\Gamma_N
\end{eqnarray}
with $\Gamma_N=\frac{2\tilde{m}m_N^2}{8\pi v^2}$ being the $N$ decay width, 
where $v= 246$ GeV as the SM vacuum expectation value. 
For other scattering processes related with leptogenesis, 
the reduced cross section are given by,
 \begin{eqnarray} \nonumber
	 \hat{\sigma_N}(x)&=&\frac{\alpha_{EW}2\pi}{xs_w^2 M_W^4}(\frac{\lambda_D v}{\sqrt{2}})^4 \\ \nonumber
	 &\times& (x
	  +\frac{2x(x-1)}{(x-1)^2+(\Gamma_N/ m_N)^2} \\ \nonumber
	  &+&\frac{x^2(x-1)^2}{2((x-1)^2+(\Gamma_N /m_N)^2)^2} \\
	  &-&(1+2\frac{(x+1)(x-1)}{(x-1)^2+(\Gamma_N /m_N)^2})\log(x+1) )  ~~~~~~~ \\
    \hat{\sigma_{Nt}}(x) &=& \frac{\alpha_{EW}2\pi}{xs_w^2M_W^4}(\frac{\lambda_D v}{\sqrt{2}})^4(\frac{x}{2(x+1)}+\frac{Log(x+1)}{x+2}) \\
    \hat{\sigma_{hs}}(x) &=& \frac{3\pi \alpha_{EW}^2M_t^2}{M_W^2s_w^2}(\frac{\lambda_D v}{\sqrt{2}})^4(\frac{x-1}{x})^2 \\ \nonumber
    \hat{\sigma_{ht}}(x) &=& \frac{3\pi \alpha_{EW}^2M_t^2}{M_W^2s_w^2}(\frac{\lambda_D v}{\sqrt{2}})^4 \\ 
	 &\times& (\frac{x-1}{x}+\frac{1}{x}Log(\frac{x-1+m_h^2/m_N^2}{m_h^2/m_N^2}))
\end{eqnarray}
where $\alpha_{EW}$ is the fine structure constant, 
$\lambda_D$ is defined by $\lambda_D=\sqrt{2\tilde{m}m_N/v^2}$, 
$s_w^2$ is the square of sine Weinberge angle, 
$M_W$ is the $W$ boson mass, 
$M_t$ is the top quark mass, 
and $m_h$ is the Higgs mass. 
Correspondingly, the reaction rate are given by,
\begin{eqnarray}
	&&\gamma_ N(z)=\frac{m_N^4}{64\pi^4 z}\int_0^{\infty}\hat{\sigma_N}(x)\sqrt{x}K_1(z\sqrt{x})dx \\
	&&\gamma_ {Nt}(z)=\frac{m_N^4}{64\pi^4 z}\int_0^{\infty}\hat{\sigma_{Nt}}(x)\sqrt{x}K_1(z\sqrt{x})dx \\
	&&\gamma_ {hs}(z)=\frac{m_N^4}{64\pi^4 z}\int_1^{\infty}\hat{\sigma_{hs}}(x)\sqrt{x}K_1(z\sqrt{x})dx \\
	&&\gamma_ {ht}(z)=\frac{m_N^4}{64\pi^4 z}\int_1^{\infty}\hat{\sigma_{ht}}(x)\sqrt{x}K_1(z\sqrt{x})dx
\end{eqnarray}

%%%%%%%%%%%%%%%%%%%%%%%%%%%%%%%%%%%%%%%%%
%%%%%%%%%%%%%%%%%%%%%%%%%%%%%%%%%%%%%%%%%
\begin{acknowledgments}
\noindent
We thank Wei Liu for the early collaboration and useful communications on this work. Hao Sun is supported by the National Natural Science Foundation of China (Grant No.12075043, No.12147205).
\end{acknowledgments}

\bibliography {qx}
\end{document}